\documentclass[superscriptaddress, prl, twocolumn]{revtex4}

\usepackage{graphicx}

\usepackage{amsfonts}
\usepackage{amsmath}
\usepackage{amssymb}

\begin{document}

\title{Comment on: Nonlocal ``Realistic'' Leggett Models Can be Considered
Refuted by the Before-Before Experiment}

\author{Marek \.Zukowski}
\affiliation{Institute of Theoretical Physics and Astrophysics,
University of Gda\'nsk, ul. Wita Stwosza 57, PL-08-952 Gda\'nsk, Poland}
%\affiliation{Tsinghua University, Beijing, China}

\begin{abstract}
It is shown here that Suarez [Found. Phys. {\bf 38}, 583 (2008)] wrongly presents the assumptions behind the Leggett's inequalities, and their modified form used by Groeblacher et al.[Nature {\bf 446}, 871 (2007)] for an experimental falsification of a certain class of non-local hidden variable models.    
\end{abstract}

\date{\today}

\maketitle

This comment is not aimed at a detailed discussion of the arguments given by Suarez in \cite{SUAREZ}. The sole aim is to clearly state that he misrepresents the assumptions behind the experiment described in \cite{Groeblacher2007}, and thus the whole set of issues associated with the Leggett's inequalities \cite{Leggett2003}. Thus the starting point of the paper is incorrect.
Therefore, the conclusions of the paper have no direct logical relation with the theory and the experimental results presented in  \cite{Groeblacher2007}. 

Suarez writes 
\begin{itemize}
\item
``Groeblacher et al. choose an explicit nonlocal dependence of Bob's outcomes on
Alice's ones, though, they note, that one can also choose any other example of a
possible non-local dependence. Thus, the local polarization measurement outcomes
$\bf A$ are predetermined by the polarization vectors $\bf u$ and an additional set of hidden
variables $\lambda$ specific to the source. The local polarization measurement outcomes $\bf B$
are predetermined by the polarization vectors $\bf u$ and $\bf v$, the set of hidden variables $\lambda$,
the settings $\bf a$ and $\bf b$, and any possible non-local dependence of Bob's outcomes on
Alice's ones. It is a crucial trait''\cite{SUAREZ}.
\end{itemize}
Let us compare the above with what is actually assumed in \cite{Groeblacher2007}.
\begin{itemize}
\item
``Let us consider a specific source, which emits pairs of photons
with well-defined polarizations $\bf u$ and $\bf v$ to laboratories of Alice and
Bob, respectively. The local polarization measurement outcomes $\bf A$
and $\bf B$ are fully determined by the polarization vector, by an additional
set of hidden variables $\lambda$ specific to the source and by any set
of parameters $\eta$ outside the source. For reasons of clarity, we choose
an explicit non-local dependence of the outcomes on the settings $\bf a$
and $\bf b$ of the measurement devices.''
\end{itemize}
 That is, no gender  asymmetry is assumed: Groeblacher et al. choose an explicit nonlocal dependence of Bob's outcomes on
Alice's local parameters {\em and Alice's outcomes on Bob's local parameters}.
This is the starting point for the derivation of the inequalities, and therefore the experiment of Groeblacher et al. pertains to this case.

Finally one should also explain that {\em only} in the Appendix I of the {\em supplementary information} for the paper (easily accessible in \cite{GR-QP})
one finds a construction of {\em an explicit toy} non-local model, which satisfies the assumptions of the form given by Suarez. But
\begin{itemize}
\item
Models satisfying these assumptions are a proper sub-class of Leggett-type models.
\item
This is just a {\em toy} model, the sole role of which is to show that the class of non-local hidden variable models introduced by Leggett contains one that  ``perfectly simulates
all quantum mechanical predictions for measurements in a plane of the Poincar\'e sphere'' \cite{GR-QP}, and therefore maximally violates the CHSH inequalities.
It {\em plays no other role} whatsoever in any other reasoning contained in the paper.
\item
The model can be trivially gender symmetrized. 
\end{itemize}

The work is a part of QAP (Qubit Applications), 6th EU Framework Programme. It has been done 
at the {\em National Centre for Quantum Information of Gdansk}. The co-authors of reference \cite{Groeblacher2007} are thanked for discussions.

\end{document}